# The Astrobiological Copernican Weak and Strong Limits for Intelligent Life

Tom Westby & Christopher J. Conselice.

University of Nottingham, School of Physics and Astronomy

## Abstract

We present a cosmic perspective on the search for life and examine the likely number of Communicating Extra-Terrestrial Intelligent civilizations (CETI) in our Galaxy by utilizing the latest astrophysical information. Our calculation involves Galactic star formation histories, metallicity distributions, and the likelihood of stars hosting Earth-like planets in their Habitable Zones, under specific assumptions which we describe as the Astrobiological Copernican Weak and Strong conditions. These assumptions are based on the one situation in which intelligent, communicative life is known to exist – on our own planet. This type of life has developed in a metal-rich environment and has taken roughly 5 Gyr to do so. We investigate the possible number of CETI based on different scenarios. At one extreme is the Weak Astrobiological Copernican principle - such that a planet forms intelligent life sometime after 5 Gyr, but not earlier. The other is the Strong Condition in which life must form between 4.5 to 5.5 Gyr, as on Earth. In the Strong Condition (under the most strict set of assumptions), we find there should be at least $36^{+175}_{-32}$ civilizations within our Galaxy: this is a lower limit, based on the assumption that the average life-time, $L$, of a communicating civilization is 100 years (since we know that our own civilization has had radio communications for this time). If spread uniformly throughout the Galaxy this would imply that the nearest CETI is at most $17000^{+33600}_{-10000}$ light-years away and most likely hosted by a low-mass M-dwarf star, likely far surpassing our ability to detect it for the foreseeable future, and making interstellar communication impossible. Furthermore, the likelihood that the host stars for this life are solar type stars is extremely small and most would have to be M-dwarfs, which may not be stable enough to host life over long timescales. We furthermore explore other scenarios and explain the likely number of CETI there are within the galaxy based on variations of our assumptions.

# 1. Introduction

One of the oldest questions that humans have asked is whether our existence – as an advanced intelligent species – is unique. While this question can be divided up into many separate problems and tangents, the main issue, in modern terminology, is whether there are other intelligent species somewhere in the visible Universe. Furthermore, this has often be framed as a question of whether there are intelligent life forms that we, in principle, could communicate with in our own Galaxy. The focus on our own Galaxy is due largely to the likely infeasibility at present at finding communication signals from more distant stellar systems, such as external galaxies.

Of course – from a statistical perspective – this is one of the most challenging problems in Science, since all we can do is attempt to learn from a single known data point (ourselves), with no possible method of modelling the distribution of the potential population of civilizations across the Galaxy. The process of this attempted extrapolation from N = 1, with no knowledge of a sample mean or standard deviation, would seem to push the integrity of logic to its limits. As Philip Ball states in Nature (2005), in his critique of the analysis of Gott (1993) on the "Implications of the Copernican Principle into our future prospects", many authors argue that "Gott had spun phantom knowledge from complete ignorance… the basic flaws lies in assigning equal probabilities to events about which we know nothing". Therefore, inevitably, the subject of extra-terrestrial intelligent and communicative civilizations will remain entirely in the domain of hypothesis until any positive detection is made, but this does not necessarily mean that we cannot propose models, based on sound logical assumptions, which may at least produce plausible estimates of the occurrence rate of such civilizations – if nothing else, we may be able to assess the likelihood of our own existence being unique, or whether the Search for Extra-Terrestrial Intelligence (SETI) is ever likely to bear fruit. Furthermore, when SETI succeeds there are implications for the uniqueness of our own civilization on Earth and our study is a reference frame for this perspective.

This issue is of monumental importance and interest to humanity but has of yet no answer – or even good guesses. There is a long history of these searches, starting with efforts by e.g., Cocconi et al. (1959), who searched for signals from extra-terrestrial intelligence without success. Searches have been greatly extended since and have been ongoing for the past few decades but still without any reliable detections, although the search area is still very small (Wright et al. 2018). Most famously Drake (1962) developed an equation which in principle can be used to calculate how many Communicating Extra-Terrestrial Intelligent (CETI: pronounced "chetee") civilizations there may be in the Galaxy. However, many of its terms are unknowable and other methods must be used to calculate the likely number of communicating civilizations.

Due to advances in astrophysics and knowledge of star formation and planetary systems we are collecting enough data to enable a new examination of the occurrence rate of CETI in the Milky Way. With new and better data on our Galaxy's star formation history and a better knowledge of the characteristics of exoplanets, we can now make a solid attempt to answer the question of the likelihood of intelligent life elsewhere. Furthermore, we argue here that we can also invert the question of how much intelligent life there is in the Universe to one in which we ask why life has not yet been found in the Galaxy, and what this implies for our own existence on Earth. The spatial distribution of intelligent lifeforms will be related to the lifetime of intelligent civilizations, including our own, thus constraining our estimate of the former will have a bearing on the latter.

We start with a revision of the Drake equation, and we make a key assumption: since the time required for the development of communicative intelligent civilization on our own planet is of order 5 Gyr, then we propose that life will have a reasonable probability of forming on another planet such as the Earth in the habitable zone of a suitable star within our Galaxy in a similar amount of time. This idea has not been confirmed but is worth exploring as on earth we see many examples of convergent evolution, and life may in principle arise in a similar manner on a different planet. In this paper we re-examine the likely occurrence rate of CETI, under two different assumptions. The first, which we call the Weak Astrobiological Copernican scenario, is that intelligent life can only form on an Earth-like planet in a habitable zone after the star is at least 5 billion years old – which mimics the amount of time it has taken to form such life on Earth (Dalrymple, 2001) – however, intelligent life can form any time after 5 billion years: in practice, this limit is not a very strong constraint as we find that most stars in the Galaxy are older than this. The other situation we investigate – called the Strong Astrobiological Copernican scenario - whereby intelligent life forms around stars exactly in the same timescale as on Earth: between 4.5 and 5.5 billion years after formation. We investigate both scenarios using the star formation history of our Galaxy, knowledge of stellar life-times and the properties of planets derived from the Kepler mission (NASA), in order to determine how many stars in our galaxy have the appropriate age to allow for the development of CETI.

This paper is a mixture of areas of contemporary astronomy, with the basic outline discussed in Section 2. Next, we discuss a variety of topics, including the star formation history of the Milky Way in Section 3. In Section 4, we investigate the metallicity distribution of stars in the Milky Way and develop a calculation of how many active CETI civilizations there are likely to be in the Galaxy using our criteria. Throughout this paper we assume a cosmology of $H_0 = 70$ kms$^{-1}$Mpc$^{-1}$, $\Omega_M = 0.3$ and $\Omega_\Lambda = 0.7$.

# 2. Estimating the Number of Intelligent Civilizations in the Galaxy

## 2.1 Background

The traditional approach towards examining whether CETI has formed in the Galaxy has been proposed through the use of the Drake equation (Drake 1962). This has remained the primary method for inferring the likely number of CETI in our Galaxy, yet it is fundamentally an unsolvable equation (prior to any extra-terrestrial life being found). This equation is nevertheless a tool for estimating the number of planets in our Galaxy that host intelligent life with the capability of releasing signals which could be detectable from Earth. It can be written as:

$$N = R_*.f_p.n_e.f_l.f_i.f_c.L$$

(Equation 1)

Where:
$N$ = the number of intelligent, communicating civilizations within the Galaxy
$R_*$ = the average star formation rate (SFR) of the Galaxy,
$f_p$ = the fraction of stars with planets
$n_e$ = the average number of planets per star that could potentially support life, per star observed to have any planets
$f_l$ = the fraction of these planets which could host life that actual develop life at some point,
$f_i$ = the fraction of these that develop intelligent life,
$f_c$ = the fraction of those which develop intelligent life that then release signals that could in principle be detected.
$L$ = the average life of an advanced civilization, or how long a civilization survives once it develops technological ability to transmit signals.

Many, but not all, of the Drake Equation terms can be simplified and calculated using new data. We have a good understanding of the star formation rate history of our Galaxy, as well as in all nearby galaxies, and the universe as a whole (e.g. Hopkins 2004, Hopkins and Beacom 2006, Bauer et al. 2011, Madau and Dickinson 2014). From Kepler data we also have a good idea concerning the fraction of stars with planets, as well as calculations of the number of these planets per star that can host life in principle.

## 2.2 The CETI Equation – a New Approach

We re-derive a modern version of a Drake-like equation by first making the simple assumption that a sufficiently Earth-like planet in the habitable zone of a suitable star which exists for a sufficiently long time (henceforth referred to as a Suitable Planet, SP) will form life in a pattern similar to what has occurred on earth (i.e. $f_l$ in Equation 1 is assumed to be 1, for an SP). This is the Astrobiological Copernican Principle. Below we give a brief overview of this idea, and how we calculate the number of CETI in our Galaxy. Later in the paper we actually make this calculation using the latest astrophysical data.

We assume that if an SP remains in the circumstellar Habitable Zone (HZ) for a time equal to the current age of the Earth (denoted as $\tau_E \approx 5$ Gyr), it will develop intelligent, communicative life. This approach has the advantage of circumventing the need for Drake equation quantities such as $f_l$ and $f_c$, which are – at present – impossible to establish on a solid, physical basis. Our assumption is based on what we call the Principle of Mediocrity: there is no evidence to assert that the Earth should be treated as a special case, and therefore – according to the Copernican Principle – we propose that the likelihood of the development of life, and even intelligent life, should be broadly uniformly distributed amongst any suitable habitats. This would also be consistent with an idea of universal convergent evolution.

It is therefore of foremost importance to estimate the fraction, $f_L$, of all stars presently within the Milky Way that are older than 5 Gyr. For this value, we take the estimate from Dalrymple (2001) to one significant figure, for – as we shall see in Section 3.1.4 - our estimated fraction is relatively insensitive to adjustments to this parameter. As we show, the results do not change significantly if we relax this criterion and allow life to form after e.g., a few Gyr, given that the star formation rate has steadily declined throughout the Galaxy's lifetime.

In the above considerations, we make the assumptions that if a planet could potentially support life, then it will inevitably develop a CETI but no earlier than $\tau_E \approx 5\ Gyr$. To determine this, we use the star formation history of the Galaxy and the Initial Mass Function (IMF) of stars. We discuss this calculation in Section 3.1 below. Clearly, our results will therefore be upper limits on the number of planets which form intelligent life. However, in the absence of data on the other terms, this is a reasonable place to begin this calculation. It is important to realize that this is in many ways the most optimistic scenario when we later discuss the number of CETI in the Milky Way we could possibly detect.

With these assumptions, we can therefore write the CETI equation as:

$$N = N_* \cdot f_L \cdot f_{sp} \cdot \left(\frac{L}{\tau'}\right)$$

(Equation 2)

Where:
$N$ = the number of intelligent, communicating civilizations in the Galaxy at the present time
$N_*$ = the total number of stars within the Galaxy
$f_L$ = the fraction of those stars which are older than 5 Gyr
$f_{sp}$ = the fraction of those stars which also host a Suitable Planet in a habitable zone, which could support life
$\tau'$ = the average amount of time that has been available in which life could have evolved on such a planet, orbiting such a star. In other words, $\tau'$ represents the time in which life could exist, which (based on our assumption) is given by:
$\tau'$ = (average age of stars in the Galaxy / Gyr) – (5 Gyr).
$L$ = the average lifetime of an advanced civilization, or how long a civilization survives once it develops a technological ability to transmit signals.

Here, the fraction $\frac{L}{\tau'}$ is of paramount importance to our estimate. In the original approach to the SETI equations (by Drake), the relevant ratio was that of the typical civilization lifetime to the entire age of the Galaxy: which, of course, assumed a constant Star Formation Rate (SFR) throughout the Milky Way's history. However, in the present work, we are concerned with the fraction $\frac{L}{\tau'}$, which can be considered as the probability of our observation of a stellar system coinciding with the (possibly relatively fleeting) existence of CETI: for example, if the average lifetime of CETI turns out to be $L \approx 200$ years, and if the average age of all stars in the Galaxy turns out to be 11 Gyr (i.e. 6 Gyr older than the critical 5 Gyr age at which we are assuming CETI can originate, hence $\tau' \approx 6$ Gyr) then the probability that we will detect CETI during its existence (which we may assume to be randomly distributed across the lifetime of the stellar system) would be $\frac{200}{6 \times 10^9} \approx 3 \times 10^{-8}$, in the Weak Astrobiological Copernican limit.

Equation 2 presents two important unknowns, L and N, which - while unknown - have well-determined lower limits of N ≥ 1 and L > 100 years, given that Earth counts as a civilization emitting radio signals and has been doing so on the order of a century. Therefore, in the most pessimistic assumption (in which we are the only intelligent communicating civilization in the Galaxy, and we are on the brink of destruction presently), N = 1 and L = 100 years. We will revisit these constraints in Section 5.

In terms of what we have referred to as a Suitable Planet, we restrict investigation to planets with a sufficiently high Earth Similarity Index (ESI), which resides within the circumstellar Habitable Zone (HZ) of a suitably old star. The fraction of stars which host such a planet – for a time sufficient to develop a communicating civilization - is referred to as $f_{HZ}$.

The possibility of a so-called Galactic Habitable Zone (GHZ) – i.e. that not all stars are able to develop life in our Galaxy due to their location – has also been considered in the past. This controversial area is debated amongst astronomers but pertains to the largely radial variation of metallicity throughout the Galaxy, as well as the density of stars, and therefore the frequency of supernovae which have the potential to destroy life once it has begun to develop. In this paper, we aim to address these issue on the grounds of the most solid physical factors, thus in Section 3.3 we tackle this by assessing the Metallicity Distribution Functions (MDFs) of stars within different regions of the Galaxy, and compute the fraction of all stars with metallicities exceeding certain selected thresholds. Therefore, we add a new term into the CETI equation to account for the fraction of stars within the Galaxy with what we shall estimate to be a sufficient metallicity for the formation of advanced biology, and - of course – to enable the existence of heavy metal resources required for a communicating civilization. We call this term $f_M$ (as set out in Section 3.3).

Hence, we can replace the term $f_{sp}$ in Equation 2 by the product $f_{sp} = f_{HZ}.f_M$ , so overall, the final form of the CETI equation is then:

$$N = N_*.f_L.f_{HZ}.f_M.\left(\frac{L}{\tau'}\right)$$

(Equation 3)

Note that the key aspects for this paper are the determination of $f_L$, L, τ′ and $f_M$; the fraction $f_{HZ}$ is based on findings from recent papers examining this fraction based on Kepler results.

Once the necessary estimates of the numerical quantities have been made, we consider twelve theoretical categories, based on different modelling assumptions which reflect different philosophical positions within the Astrobiological Copernican Principle. This Principle asserts that the properties and evolutionary mechanisms in operation in our Solar System is not unusual in any important way, and so we may feel justified in assuming that life, and even communicative intelligence, should stand an equal chance of evolving in any such system, given the requisite amount of time and raw materials. Our twelve modelling categories are as shown in Table 1:

| Category | Comment about A.C.P. | Assumption 1: concerning the time interval available for the existence of life. | | Assumption 2: minimum stellar metallicity required for CETI. |
|---|---|---|---|---|
| 1 | Ultra-Weak | (Primitive Life only). Assume that Primitive Life becomes established rapidly, wherever suitable, stable conditions arise, and will persist for the entire stellar lifetime. | | $0.1Z_\odot$ |
| 2 | | | | $0.5Z_\odot$ |
| 3 | | | | $1.0Z_\odot$ |
| | | CETI possible in stellar system of age: | Value of $\tau'/Gyr$ implied by Assumption 1. | |
| 4 | Weak | (age / Gyr) > 5.0 | (Average stellar age / Gyr) − (5.0 / Gyr) | $0.1Z_\odot$ |
| 5 | | | | $0.5Z_\odot$ |
| 6 | | | | $1.0Z_\odot$ |
| 7 | Moderate | 4.0 < (age / Gyr) < 6.0 | 2.0 Gyr | $0.1Z_\odot$ |
| 8 | | | | $0.5Z_\odot$ |
| 9 | | | | $1.0Z_\odot$ |
| 10 | Strong | 4.5 < (age / Gyr) < 5.5 | 1.0 Gyr | $0.1Z_\odot$ |
| 11 | | | | $0.5Z_\odot$ |
| 12 | | | | $1.0Z_\odot$ |

Table 1: Describing the twelve categories of differing modelling assumptions, relating to different relative strengths of the Astrobiological Copernican Principle (A.C.P.)

Note: In the Ultra-Weak case (Categories 1, 2 and 3) the fraction $\left(\frac{L}{\tau'}\right)$ in Equation 3 is set to 1, and the term $f_L$ is set to 1.

# 3. Calculations

## 3.1 Calculation of $f_L$: the fraction of stars which exist in the Galaxy today, and which are older than 5 Gyr.

### 3.1.1 Star Formation Rate (SFR) History.

The first parameter we investigate is $f_L$ – the fraction of stars within the Milky Way Galaxy which are older than 5 Gyr. This relates to our assumption that communicating intelligent life can form after this time period, which we are making based on the fact that intelligent life on the Earth took approximately this long to developed (see Dalrymple, 2001 – note: we have taken the best estimate for the age of the Earth as $4.54^{+0.05}_{-0.05}\ Gyr$, and we express this to one significant figure). Clearly, this is the only time-scale we have as an example for the formation of intelligent life, and it is, perforce, a simplified first approximation: we are considering the persistence of the stability of the star's conditions, and can say nothing about the planetary environment, which may be dramatically affected by climate, orbital or geological shifts within this 5 Gyr timeframe. In fact, because of this our limits are, in many ways, upper limits due to these other conditions. However, even if we relax this assumption, we find very little difference in the following results (see Section 3.1.4), and we demonstrate that – even within this apparently severe limit - we obtain interesting results.

To carry out this calculation we need to determine the age distribution of stars within our Galaxy. To do this, we assume a form of the star formation rate history and then convert this into the number of stars formed at a given mass as a function of time, throughout the entire history of the Galaxy. To do this first part, we use an analytical fit for data on Star Formation Rate (SFR) versus redshift (z).

What we want to use here is a function that describes the variation in SFR within the Milky Way Galaxy throughout time, however, this is difficult to know and no functional formula or derivation of this exists. We start by using established SFR data for distant galaxies throughout the Universe at large, as reported by Madau and Dickinson (2014): the data from Table 1 of that paper – showing $\log(SFR/M_\odot\ year^{-1}\ Mpc^{-3})$, together with their associated error bars, versus redshift, z - is used in the construction of Figure 1, below. This is likely a good presentation of past SFR within the Milky Way as the star formation history of the Local Group matches the global history fairly well (e.g. Weisz et al. 2014). We do not have the exact values of the Milky Way's star formation history, so we use knowledge of the shape of the global history and renormalize this based on the known volume of our Galaxy and the number of stars it contains today.

To do this we take the analytical form of the SFR history of the universe and adapt it as the relative star formation history of the Milky Way, which is the relevant quantity for this work. For this we use the variation of SFR with redshift which can be fitted as an analytical expression, as shown by many authors (e.g., Hopkins, 2004; Hopkins and Beacom, 2006; Hernquist et al. 2003). In this last work, observed data on SFR at different redshifts, z (representing different times in cosmic history) is fitted with a function of the following form:

$$\dot{\rho}_*(z) = \dot{\rho}_*(0) \frac{\chi^{n_1}}{1 + \alpha(\chi - 1)^{n_2} exp(\beta \chi^{n_3})}$$

(Equation 4)

Where:

$\dot{\rho}_*(z)$ = Star Formation Rate, in units $M_\odot Mpc^{-3} yr^{-1}$, as a function of redshift,

$\dot{\rho}_*(0)$ = Present-day value of SFR

and $\chi$ is a function of redshift, (z), which is defined by:

$$\chi(z) = \left(\frac{H(z)}{H_0}\right)^{\frac{2}{3}}$$

$H(z)$ represents the Hubble parameter, as a function of redshift, in units $kms^{-1} Mpc^{-1}$, which is defined by:

$$H(z) = H_0[(\Omega_M(1+z)^3 + (1 - \Omega_M - \Omega_\Lambda)(1+z)^2 + \Omega_\Lambda)]^{\frac{1}{2}}$$

Here, $H_0$ is the present day value of the Hubble constant, taken as 70 $kms^{-1} Mpc^{-1}$, and $\Omega_M$ (the density parameter of matter), and $\Omega_\Lambda$ (the density parameter of Dark Energy) are taken as 0.3 and 0.7 respectively in this work.

$\alpha, \beta, n_1, n_2$ and $n_3$ are the constants whose values may be varied to achieve the best fit to the observed data for SFR vs z. Curve-fitting techniques yield the values of the parameters which give a best fit to the observational data. This fit yields the following values for the fitting constants in Equation 4, which are shown in Table 2:

| Fitting Constant | Value for fitting constant (with associated range, where possible) derived from curve-fitting analysis | Range of values of $f_L$ obtained from Fig 3, based on the range in this fitting constant |
|---|---|---|
| $\alpha$ | $0.528^{+1.472}_{-0.438}$ | $0.967 < f_L < 0.973$ |
| $\beta$ | $2.36^{+0.64}_{-0.96}$ | $0.926 < f_L < 0.996$ |
| $\dot{\rho}_*(0)$ | $0.330^{+1.670}_{-0.230}$ | $0.968 < f_L < 0.968$ |
| $n_1$ | $5.91^{+1.47}_{-1.91}$ | $0.906 < f_L < 0.997$ |
| $n_2$ | $-0.3508$ (no range – see below) | $0.968$ (no range – see below) |
| $n_3$ | $1.22^{+0.68}_{-0.42}$ | $0.782 < f_L < 0.999$ |

Table 2: Fitting Constants for Equation 4. Note, the curve-fitting analysis was highly volatile to slight changes in the parameter $n_2$, so this was not varied to allow for an exploration of the range of the other fitting constants.

We then combine this relationship of the SFR with redshift with the relationship between redshift and lookback time, $t_L$ (i.e. the time between the emission of the light from a distant source, and the present time at which the light is received by the observer). We then plot the cosmic history of star formation: see Figure 1, in which the raw SFR data has been combined with the analytical fitting function Equation 4, using parameters from Table 2. Note that the raw data has associated error bars, and we have shifted some points slightly to avoid overlap. Note also that in Fig 1, the green curve shows the fitting function in which the central values of the fitting constants are employed.

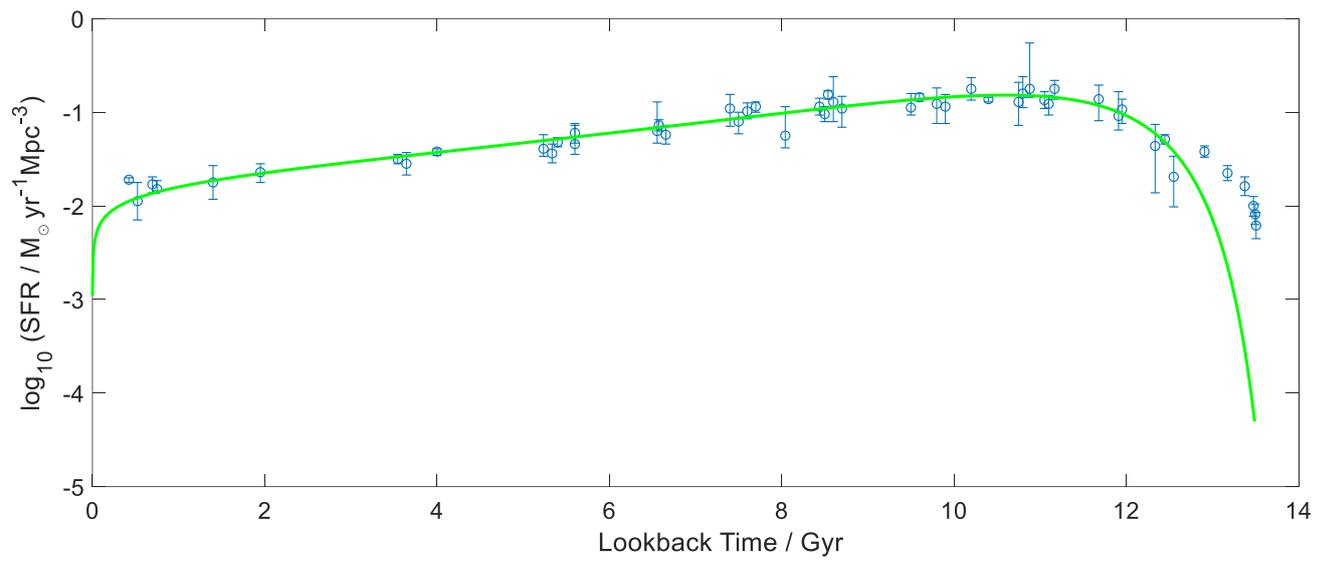

**Fig 1: Cosmic Star Formation History Data (from Madau and Dickinson, 2014) with analytical fit using Eqn 4**

## 3.1.2 Stellar Mass Distribution and the Main Sequence Lifetimes of Stars.

### i) The Distribution of Stellar Masses according to the Salpeter Initial Mass Function (IMF)

At this point, we must introduce the Initial Mass Function (IMF), in order to calculate the distribution of the numbers of stars, N, with masses, M, between certain mass limits ($M_{lower}$ and $M_{upper}$). The starting point we consider here is the Salpeter IMF (Salpeter 1955), described in the following way:

$$\frac{dN}{dM} = kM^\alpha \qquad \text{(Equation 5)}$$

where k is a normalization constant, and α = − 2.35. Hence:

$$\int dN = k . \int_{M_{lower}}^{M_{upper}} M^\alpha dM$$

and therefore:

$$N = \frac{k}{\alpha+1} . [M^{\alpha+1}]_{M_{lower}}^{M_{upper}} \qquad \text{(Equation 6)}$$

If we normalize this expression such that N = 1, this allows us to find the relative fraction of stars in the entire stellar population which formed with masses between any desired limits. To do this, we evaluate the normalization constant k, by using values for the minimum and maximum possible stellar mass as $M_{lower}$ and $M_{upper}$ respectively.

The minimum mass required for Hydrogen fusion is taken as 0.08$M_\odot$ (Richer et al., 2006 – note: this choice of lower mass limit may have an impact on the final value for $f_L$, see Section 3.1.4); whilst the maximum stellar mass we use is 100$M_\odot$ (Kroupa, 2005) - there is considerable debate on this upper limit, but fortunately this value will prove far less significant at the higher end of the Salpeter distribution.

Using these values, we set the left-hand side of Equation 6 equal to unity, in order to achieve an expression for the relative fraction of all stars between the given mass limits. Therefore, we evaluate k = 0.0446. Hence, we can re-write Equation 6:

$$\frac{N}{N_{total}} = 0.033 \times [M^{\alpha+1}]_{M_{upper}}^{M_{lower}} \qquad \text{(Equation 7)}$$

which we use to evaluate the relative fraction of stars in a population with masses between the desired limits. We explore the relationship between this Survival Fraction and the time since starburst and find that the vast majority of the stars which still survive today did indeed form in starbursts more than 5 Gyr ago.

## ii) Stellar Masses and Lifetimes.

We now consider the Main Sequence lifetime of stars, in order to ascertain the fraction of stars which formed at a certain time in the past and which still survive today. The first step here is to establish the relationship between stellar mass and a star's lifetime on the Main Sequence. The amount of time that a star spends burning Hydrogen will, of course, depend on its initial mass and luminosity, as: $\tau_{MS} \sim \left(\frac{M}{L}\right)$. Given that the estimated Main Sequence lifetime of the Sun is of order 10 Gyr (Schroder and Smith, 2008), we may write, in terms of solar units:

$$\frac{\tau_{MS}}{Gyr} \sim 10 \left(\frac{M}{M_\odot}\right)\left(\frac{L}{L_\odot}\right)^{-1} \qquad \text{(Equation 8)}$$

where $L$ is the luminosity of the star. Luminosity is also tightly dependent on stellar mass, and – taking the relationships from Salaris et al. (2005):

$$\frac{L_{MS}}{L_\odot} \approx 1.4 \left(\frac{M}{M_\odot}\right)^{3.5} \quad \text{for } 2\ M_\odot \leq M \leq 20\ M_\odot$$

$$\frac{L_{MS}}{L_\odot} \approx \left(\frac{M}{M_\odot}\right)^{4} \quad \text{for } 0.43\ M_\odot \leq M < 2\ M_\odot$$

$$\frac{L_{MS}}{L_\odot} \approx 0.23 \left(\frac{M}{M_\odot}\right)^{2.3} \quad \text{for } M < 0.43\ M_\odot$$

(Equations 9)

Hence, combining the relationships from Equations 8 and 9, we find:

$$\tau_{MS}/Gyr \approx 7.1 \left(\frac{M}{M_\odot}\right)^{-2.5} \quad \text{for } 2\ M_\odot \leq M \leq 20\ M_\odot$$

$$\tau_{MS}/Gyr \approx 10 \left(\frac{M}{M_\odot}\right)^{-3} \quad \text{for } 0.43\ M_\odot \leq M < 2\ M_\odot$$

$$\tau_{MS}/Gyr \approx 43 \left(\frac{M}{M_\odot}\right)^{-1.3} \quad \text{for } M < 0.43\ M_\odot$$

(Equations 10)

Since we are interested in evaluating the fraction of all stars which are older than 5 Gyr, we examine only stars of spectral types F, G, K, M and lower (if we include L and T-type dwarfs in our consideration), for which we take the mass data in Table 3:

| Spectral Type | F | G | K | M |
|---|---|---|---|---|
| Mass, M / M$_\odot$ | 1.04 < M < 1.4 | 0.8 < M < 1.04 | 0.45 < M < 0.8 | 0.08 < M < 0.45 |
| Proportion of all stars belonging to this Type (calculated from Eqn 7) | 1.03 % | 1.33 % | 5.24 % | 90.15 % |

Table 3: Stellar Masses of spectral type F, G, K and M, from Habets and Heintze (1981)

For example, a stellar mass value of 1.26 M$_\odot$ corresponds to a Main Sequence lifetime of 5 Gyr, which is the critical value we use in our determination of $f_L$.

One conclusion from our results is that the most likely location for CETI life is around low-mass M-dwarf stars, since we estimate that these contribute around 90 % of the total population. It is thus important to note that one major issue in Astrobiology and the development of life outside the Earth is the lack of understanding as to whether M dwarf stars are likely hosts for the stable conditions required for the development of intelligent, or even basic, life. This is a particular problem for the present analysis, since the numbers of low-mass, long-lived stars will be dominated by these M dwarfs. This is an issue of on-going debate, and new exoplanet discoveries in M dwarf systems (such as Proxima Centauri and Trappist-1) raise new questions about the planetary environments, especially for small planets in close, tidally locked orbits, around volatile, small stars. Wandel (2018) presents a contemporary discussion on these issues, and Haqq-Misra, Kopparapu and Wolf (2017) consider the implications of our existence around a (less abundant) yellow star, as opposed to a (more abundant) red star, in light of the Copernican Principle. It is hoped that future exoplanet discoveries will help to refine the relative abundance of stable planetary environments around M-dwarf stars.

## 3.1.3 The Distribution of Ages of Stars Surviving in the Galaxy.

In this section we investigate the survival fraction of stars in the Milky Way over time and use this to arrive at the age distribution of all stars surviving in the Galaxy today. First, we use the analytical fit to the raw SFR data to calculate the total mass of stars formed per Mpc$^3$, per 50Myr interval of time, as a function of the time since each star-forming event.

However, more significant for our purpose is the mass of stars per Mpc$^3$ that was formed during this 50Myr time step, and still survives today. We then renormalize this such that the integral of the number of these stars is equal to the total number we know exist today in the Milky Way. To achieve this, we use the Salpeter IMF to calculate the distribution of the total mass of stars made up by stars of each mass. The number of stars formed in a particular mass category, N, is given by Equation 6. Therefore, the total mass of the stars within each mass category is the number of stars of a given mass multiplied by that mass, giving:

$$M_{total} = b . \int_{M_{lower}}^{M_{upper}} M . M^{\alpha} . dM$$

$$M_{total} = \frac{b}{\alpha + 2} . [M^{\alpha+2}]_{M_{lower}}^{M_{upper}}$$

(Equation 11)

where b is another normalization constant, and again, α = -2.35 for a Salpeter IMF. We normalize this expression to make $M_{total}$ = 1 for the full range of masses, from $M_{lower}$ = 0.08M$_\odot$ to $M_{upper}$ = 100M$_\odot$, giving:

$$\frac{M}{M_{total}} = 0.450 \times [M^{-0.35}]_{M_{upper}}^{M_{lower}}$$

(Equation 12)

which can be used to find the relative fraction of the grand total of stellar mass which was formed in a particular category of stars of masses between the desired limits. (Note: once more, the evaluation of the normalization constant in Equation 12 is dependent on the choice of minimum mass of 0.08M$_\odot$ - see discussion in 3.1.4).

In Fig 2, we plot the total mass of stars per Mpc$^3$, which formed in a 50Myr time-step at the time shown, $M_{formed}$: this is done by multiplying the Star Formation Rate (SFR) at a given time, $\dot{\rho}_*$, by our time-step $50 \times 10^6 \, yr$:

$$M_{formed} = \dot{\rho}_* \times 50 \times 10^6. \qquad \text{(Equation 13)}$$

The resulting values are shown as the top curve in Fig 2 (the solid red line). This curve shows that – as previously stated – the peak in SFR occurred at a time approximately 10.5 Gyr ago, when the Universe was approximately 3.3 Gyr old.

The second curve from the top in Fig 2 (red, dashed line) shows the total mass of stars, per Mpc$^3$ which still survives today from each time-step. This is formulated in the following way: first, the relationships in Equations 10 are used to express stellar mass as a function of Main Sequence lifetime, $M(\tau_{MS})$, then this is used in Equation 12 to arrive at an expression for the fraction of stellar mass formed during starbursts at some time in the past and which still survives today, $f_{survived}(L)$, as a function of this lookback time. Multiplying the function shown by the top curve by this survival fraction, we obtain the total mass of stars which still survive today from the starburst at that time in the past,

$$M_{survived} = f_{survived}(L) \times M_{formed} \qquad \text{(Equation 14)}$$

– this is shown by the second curve (red, dashed line). Hence – for instance – at a time 5 Gyr in the past, some 2.44 x 10$^6$ M$_\odot$ Mpc$^{-3}$ of stellar mass was formed during a 50 Myr time-step, of which some 1.51 x 10$^6$ M$_\odot$ Mpc$^{-3}$ survives today: a mass survival fraction of 62%.

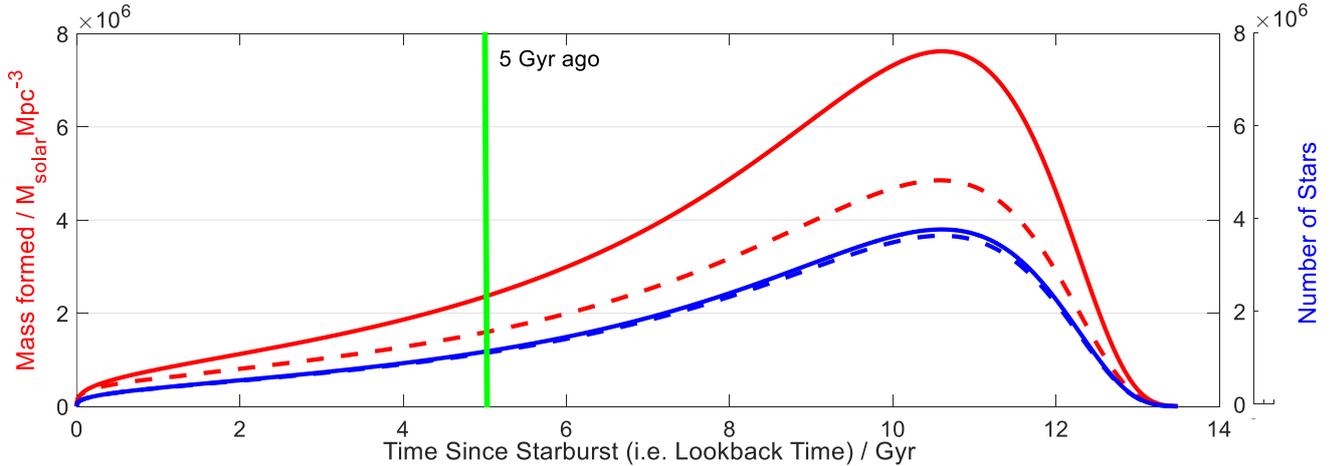

**Fig 2: Top plots (RED): total mass of stars formed (SOLID line) and surviving today (DASHED line) per cubic-Mpc, in 50 Myr time-steps.**

**Bottom plots (BLUE): total number of stars formed (SOLID line) and surviving today (DASHED line) per cubic-Mpc, in 50 Myr time-steps.**

**All plots normalized to ensure total number of stars in Galaxy today, N$_*$ = 2.5 x 10$^{11}$, and volume of Galaxy = 226 kpc$^3$**

If we again employ the Salpeter IMF for the total stellar mass, but use $M_{formed}$ to serve as the $M_{total}$ in Equation 11, we can evaluate a new value for the normalization constant, b. This will then be worked back into the Salpeter IMF expression (Equation 6), allowing the number of stars formed per Mpc$^3$ per 50 Myr time-step to be generated. The function expressing the fraction of stars which have survived since a starburst at a certain lookback time is then employed to create an accompanying plot of the number of these stars (which were formed during a particular 50 Myr time-step) which still survive today. The resulting plots are shown in Fig 2 as the two lower curves (blue): the blue solid line represents total number of stars formed per Mpc$^3$ per 50 Myr time-step, whilst the blue dashed line represents the number of those stars which still survive today.

Note that, in order to create the blue plots in Fig 2, it was necessary to properly normalize the functions used in the preceding, red plots, which were based on SFR data averaged over the Universe, and we cannot be sure that this represents the SFR within the Milky Way fairly. Hence, we evaluated the area under the blue curves, and scaled the functions to properly represent the total number of stars in the Milky Way (over which there is a considerable range in estimated values, and we adopt an approximation of 250 billion stars – see Masetti 2015) and the approximate volume of the Galaxy (which we model as a uniform cylindrical thin disk of volume 226 kpc$^3$ (see Rix 2013).

For example, Fig 2 shows that – during a 50Myr time step, at a time 5 Gyr ago – a total of 1.26 x 10$^6$ stars were formed throughout the Milky Way Galaxy as a whole, of which 97% survive today. In fact, we find that most stars formed still exist today, even if a larger fraction of stellar mass has been recycled. This is due to our assumption of the Salpeter IMF.

Note also that the red and blue curves in Fig 2 demonstrate an important point about the calculation of $f_L$: the total mass of stars formed in starbursts in the past has decreased today by a substantial fraction, but – since the vast majority of stars are low-mass, and these have the greatest Main Sequence lifetimes – the actual number of stars which still survive today has decreased by a small amount. Therefore, the fraction of stars surviving today which are older than 5 Gyr, $f_L$, should be close to 100%. However, this calculation also allows us to determine the average age of the stars, which we use in Section 3.2.

As a final stage in this section, we create a plot showing the age distribution of all stars in the Galaxy today, (i.e. a plot of the cumulative number of stars surviving until today, vs. the time since their formation) by performing a numerical integral of the bottom (blue dashed) plot in Fig 2, using the 50 Myr time-steps. The resulting cumulative plot is shown in Fig 3.

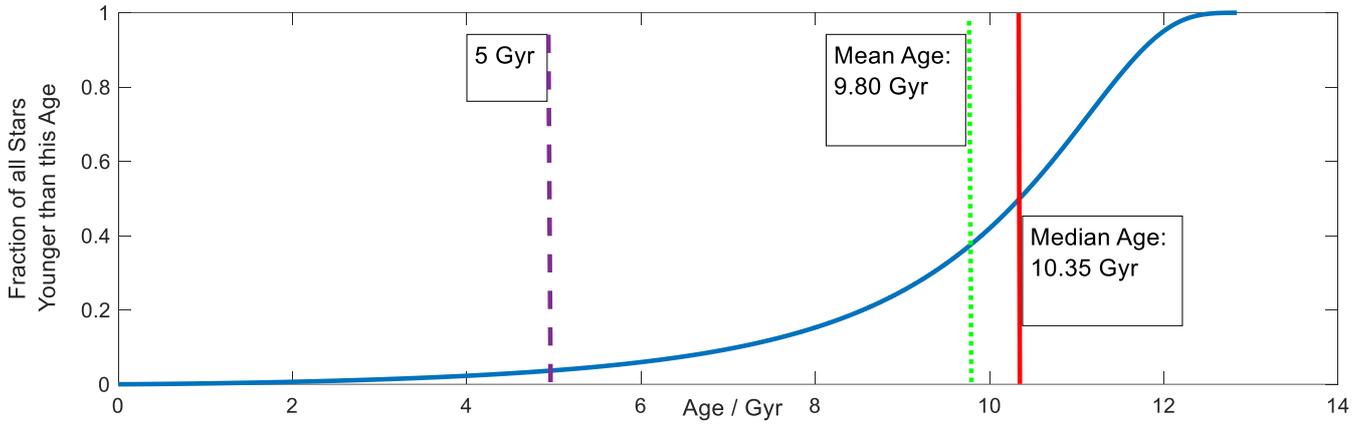

**Fig 3: Age Distribution of All Surviving Stars in the Milky Way Galaxy today.**
**Vertical lines shown at age = 5 Gyr (dashed purple line) - used in the determination of $f_L$**
**and the mean age (dotted green line) and median (solid red) - both used in Section 3.2**

This figure shows that the fraction of all stars surviving in the Milky Way today which are older than 5.0 Gyr: $f_L = 0.963^{+0.034}_{-0.183}$

This estimate is based on using the central values of the fitting constants, given in Table 2. This table also presents the corresponding values of $f_L$ which come from the use of the upper and lower values of the fitting constants, allowing for an estimate of the uncertainty in $f_L$ to be made.

## 3.1.4 Choice of I.M.F. and Uncertainty in the $f_L$ value.

In this section, we calculate the value of $f_L$ by an independent method using the Chabrier IMF for comparison. We consider the Chabrier IMF for individual (non-multiple) stars (Chabrier 2003), described in the following way:

$$\frac{dN}{dM} = 0.158 \left(\frac{1}{\ln(10M)}\right) \exp\left(-\left(\frac{(\log(M)-\log(0.08))^2}{2 \times 0.69^2}\right)\right) \quad for \ M < 1 M_\odot$$

$$\frac{dN}{dM} = km^{-\gamma} \quad for \ M > 1 M_\odot$$

(Equations 15)

where γ = 2.3 ± 0.3, and k is chosen to provide a smooth transition between the two regions.

According to our calculation, the value of $f_L$, based on the Chabrier IMF, is approximately: $f_{L,Chabrier} \approx 97\%$. Hence, we conclude that the choice of the Initial Mass Function has a low impact on our overall accuracy in $f_L$. This is due to the number of stars being dominated by the lowest mass systems – the M dwarfs in particular.

It is worth discussing some of the implications concerning the result that 97% of the stars in the Milky Way are older than 5 Gyr. Firstly, we find that the somewhat arbitrary assumption that CETI becomes established when the star is 5 Gyr old will not have a significant impact on our final result on the number of civilizations within the Galaxy, since there are relatively few stars younger than 5 Gyr: for instance, our calculation based on the critical time of 4.5 Gyr yields a value of $f_L \approx 97.4\%$, whereas – for 5.0 Gyr, we find of $f_L \approx 96.3\%$. Upon repeated calculation, as we vary this assumption of the time required for life to be established over the range 3.0 Gyr to 5.0 Gyr, we find a value of $f_L = 97.5\%^{+1.2\%}_{-1.2\%}$, showing only small deviation from the estimate of 97%, which is based on the 5 Gyr assumption. This part of our calculation shows that the vast majority of stars in the Milky Way are in principle old enough to develop life as has occurred on Earth. The Solar System formed late in the history of the Galaxy and most stars within the Milky Way are older than it.

The choice of minimum stellar mass at $0.08 M_\odot$ - taken as the minimum mass required for Hydrogen fusion – may be expected to have a greater impact, given the large abundance of stars at the lower end of the mass range. However, by modifying the assumption of minimum stellar mass (over the range $0.06 M_\odot$ to $0.10 M_\odot$), we recalculate $f_L$, and our model appears to be quite insensitive to the choice of minimum stellar mass:

$f_L = 96.27\%^{+0.02\%}_{-0.02\%}$.

## 3.2 Calculation of Technical Civilization Evolution time, τ': the average time available for CETI.

Next we calculate the value of τ' used in Equations 2 and 3, representing the average length of time that a star in the Galaxy has spent beyond the age of 5 Gyr (which is our assumed time at which a communicating intelligent civilization can become established). Hence, τ' represents the average time available for the existence of intelligent civilizations around a star. Since the vast majority of stars (~97%) are older than 5 Gyr, then, to a very close approximation, we can say:

τ' = (Average age of stars in Galaxy / Gyr) – 5 Gyr          (Equation 16)

Fig 3 shows the cumulative number of stars in the Galaxy distributed with stellar age. We compute an average age of a Milky Way star, by firstly, multiplying the number of stars of age t, $(n_t)$ surviving today (which comes from the bottom dashed blue plot of Fig 2), by their average age, t. An average stellar age can be processed by finding a cumulative plot of this data. This process yields a mean age for the stars within the Milky Way as:

$$mean\ age = \frac{\sum t.n_t}{N_*} = \frac{2.451 \times 10^{12}\ Gyr}{2.5 \times 10^{11}} = 9.80\ Gyr \qquad (Equation\ 17)$$

Note: in this method, we have worked out the mean age of all stars in the Galaxy. However, we require the average age of all of the stars which have survived beyond 5 Gyr. But since the fraction of stars older than 5 Gyr, $f_L$, is found from Section 3 to be so large (97%), the difference will be minimal. We calculate that the average age of all of the stars which have survived beyond 5 Gyr is 9.80 Gyr (which matches the 'Lookback time' of the known peak in star formation rates, at redshift, z ≈ 2). Hence, the value of τ' in (Equation 2 and 3) is

$$\tau'_{mean} = (9.80 - 5)\ Gyr = 4.80\ Gyr$$

This estimate of τ' can be compared to an independent estimate of the median age, based on the cumulative age distribution of all stars in the Galaxy, shown in Fig 3. This process yields a median age for the stars within the Galaxy as: 10.35 Gyr. Hence:

$$\tau'_{median} = (10.35 - 5)\ Gyr = 5.35\ Gyr$$

If we use these two independent estimates as the basis for our uncertainty value, we find:

$$\tau' = 4.80 \pm 0.55\ Gyr \qquad \text{which gives a percentage uncertainty in τ' of 11%}$$

## 3.3 The fraction of stars with sufficient metallicity for advanced life ($f_M$)

### 3.3.1 Metallicity Distribution Functions (MDFs) as a function of position in the Galaxy.

Having developed an estimate of the fraction of all stars in the Milky Way which are older than 5 Gyr, it now remains to consider the issue of their metallicity, Z. Being older than 5 Gyr does not necessarily mean that a star is a likely target in which to search for life, as it may be a very old Population II star with low metallicity, which could presumably rule out the presence of rocky planets and lifeforms. What is required is an investigation into the Metallicity Distribution Functions (MDFs) of stars throughout the Galaxy, so that we may evaluate an estimate of $f_M$: that is to say, the fraction of all Milky Way stars with a metallicity greater than some reference value. Using this as part of our criteria is an important aspect and again relates to the Copernican principle: life on Earth has formed in a very metal-rich environment, thus it is seems likely that life would normally form in a metal-rich environment on other planets. One could argue – a priori - that having a stellar environment with metallicity which exceeds a certain reference value could be a prerequisite for the formation of habitable planets and even life itself. This is an assumption we make hereafter, but later discuss other possibilities.

According to Johnson and Li (2012), a suitable minimum stellar metallicity required for the formation of planets with Earth-like characteristics has been posited as 0.1 $Z_\odot$. However, for the present work, we employ three reference values in the investigation: 0.1$Z_\odot$, 0.5$Z_\odot$ and 1.0$Z_\odot$ and explore how the results would vary within these assumptions.

Hayden et al (2015) present details of their investigations into MDFs throughout different regions of the Milky Way disk, in the form of skewed Gaussian distributions, with particular values of mean, standard deviation and skewness, as recalculated for our purposes in Table 6-8, below. These are used to generate a distribution, and then the required percentages of stars above the three reference values of metallicity are determined.

Hence – for example – by using these measurements, it is possible to arrive at the following estimates for the Galactic region which contains the Sun (which we refer to as Region D – see Table 6):

Fraction of stars in Region D with metallicity greater than the reference value $1.0\ Z_\odot$:
$f_{M,1.0} = 0.5030^{+0.0025}_{-0.0025}$

Fraction of stars in Region D with metallicity greater than the reference value $0.5\ Z_\odot$:
$f_{M,0.5} = 0.9360^{+0.0007}_{-0.0007}$

Fraction of stars in Region D with metallicity greater than the reference value $0.1\ Z_\odot$:
$f_{M,0.1} > 0.9999$

This process has been carried out for each of the MDFs detailed in Hayden et al (2015), and in Table 4, below, and Tables 13 and 14, in Appendix 1, we show the calculated percentages of stars within each region of the Galaxy with a metallicity lower than the reference values 1.0 $Z_\odot$, 0.5 $Z_\odot$ and 0.1 $Z_\odot$ respectively. The stated uncertainties in the skewness value in each table is used to generate minimum and maximum estimates of these calculated percentages.

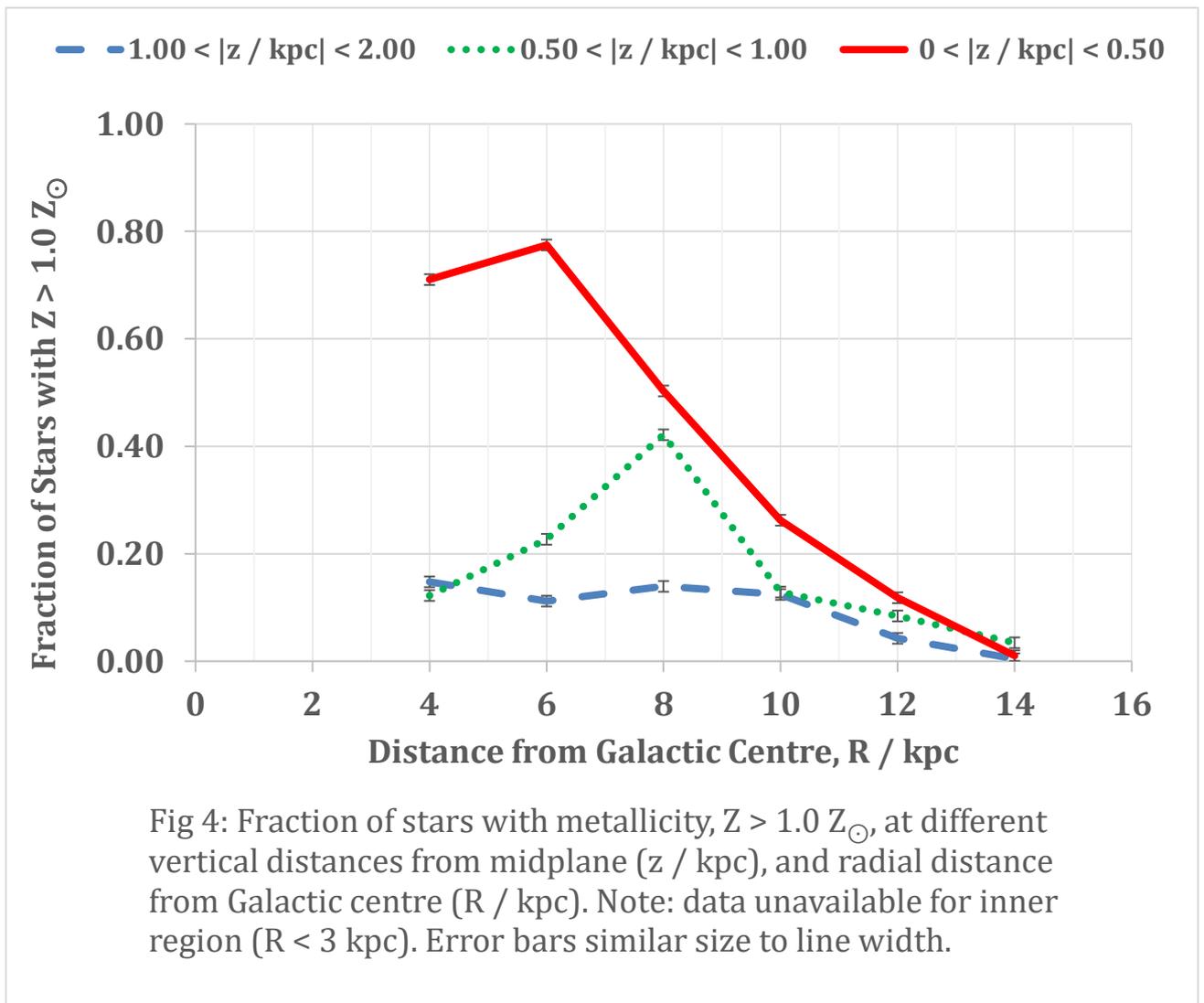

Fig 4: Fraction of stars with metallicity, Z > 1.0 $Z_\odot$, at different vertical distances from midplane (z / kpc), and radial distance from Galactic centre (R / kpc). Note: data unavailable for inner region (R < 3 kpc). Error bars similar size to line width.

| Region | Distance to Galactic centre, R / kpc | Mean Metallicity / dex | Standard Deviation / dex | Skewness (together with its uncertainty) | Calculated Percentages of stars in this region with Z < reference value: 1.0 $Z_\odot$ |
|---|---|---|---|---|---|
| | Distance from mid-plane, $|z|$ / kpc:  0.00 < $|z|$ < 0.50 | | | | |
| B | 3 < R < 5 | +0.23 | 0.24 | −1.68 ±0.12 | $28.97^{+0.80}_{-0.79}$ % |
| C | 5 < R < 7 | +0.23 | 0.22 | −1.26 ±0.08 | $22.52^{+0.44}_{-0.43}$ % |
| D * | 7 < R < 9 | +0.02 | 0.20 | −0.53 ±0.04 | $49.70^{+0.25}_{-0.25}$ % |
| E | 9 < R < 11 | −0.12 | 0.19 | −0.02 ±0.03 | $73.76^{+0.14}_{-0.15}$ % |
| F | 11 < R < 13 | −0.23 | 0.19 | +0.17 ±0.06 | $88.19^{+0.19}_{-0.18}$ % |
| G | 13 < R < 15 | −0.43 | 0.18 | +0.47 ±0.13 | $98.97^{+0.06}_{-0.06}$ % |
| | Distance from mid-plane, $|z|$ / kpc: 0.50 < $|z|$ < 1.00 | | | | |
| I | 3 < R < 5 | −0.33 | 0.32 | −0.50 ±0.11 | $87.77^{+0.48}_{-0.46}$ % |
| J | 5 < R < 7 | −0.18 | 0.29 | −0.50 ±0.09 | $77.31^{+0.54}_{-0.56}$ % |
| K | 7 < R < 9 | −0.02 | 0.25 | −0.49 ±0.06 | $57.86^{+0.43}_{-0.44}$ % |
| L | 9 < R < 11 | −0.23 | 0.21 | −0.22 ±0.10 | $87.12^{+0.33}_{-0.34}$ % |
| M | 11 < R < 13 | −0.27 | 0.19 | +0.28 ±0.11 | $91.58^{+0.27}_{-0.28}$ % |
| N | 13 < R < 15 | −0.33 | 0.19 | −0.60 ±0.39 | $96.57^{+0.34}_{-0.42}$ % |
| | Distance from mid-plane, $|z|$ / kpc: 1.00 < $|z|$ < 2.00 | | | | |
| P | 3 < R < 5 | −0.27 | 0.29 | −0.48 ±0.14 | $85.23^{+0.62}_{-0.67}$ % |
| Q | 5 < R < 7 | −0.33 | 0.29 | −0.32 ±0.13 | $88.81^{+0.49}_{-0.53}$ % |
| R | 7 < R < 9 | −0.27 | 0.28 | −0.53 ±0.06 | $86.07^{+0.25}_{-0.26}$ % |
| S | 9 < R < 11 | −0.27 | 0.25 | −0.37 ±0.13 | $87.58^{+0.47}_{-0.50}$ % |
| T | 11 < R < 13 | −0.38 | 0.23 | −0.40 ±0.21 | $95.74^{+0.29}_{-0.32}$ % |
| U | 13 < R < 15 | −0.43 | 0.17 | −0.60 ±0.73 | $99.54^{+0.08}_{-0.14}$ % |

Table 4: Calculated Percentages of stars with metallicity lower than the reference value of 1.0 $Z_\odot$, within each region of the Galaxy. (Note: Region D, containing the Sun, is shown with the asterisk * in bold).

# 3.3.2 Percentages of Stars throughout the whole Galaxy which exceed the Metallicity Reference Values.

In order to calculate an estimate of the overall percentage of stars in the Galaxy with metallicities exceeding the Reference Values, we require an estimate of the proportion of all stars which are located within the different regions analysed in the last section. To express this, we create simple exponential models for the decrease in number density of stars throughout the Galaxy, with increasing Galactic radius and increasing vertical distance from the mid-plane: $n(R)$ and $n(z)$ respectively.

We model the variations in stellar number density with the following functions:

$$n(R) \alpha \, e^{(-\frac{R}{h_R})}$$

$$n(z) \alpha \, e^{(-\frac{z}{h_z})} \qquad \text{(Equations 18)}$$

where we can take the scale length, $h_R$, and scale height, $h_z$, from Bland-Hawthorne and Gerhard (2017):

$h_R$: 2.5 ± 0.4 kpc
$h_z$: between 220 pc and 450 pc (with a mean estimate of 335 pc).

The number density modelling functions in Equations 18 are normalized and plotted, and the percentages of all stars within each of the regions mentioned in Table 4 are estimated from measurements on these plots.

The results for the fraction of the number density of stars within each region of the Galaxy (according to the exponential decay models of number density, Equations 18) are shown in Tables 5 and 6. Error bounds are generated from further plots of the functions which acknowledge the uncertainty range in the exponential scale-length and scale-height. Note, in order to create fairly weighted fractions of the absolute numbers of stars within each region (given in Table 11), we need to take into account the fact that these regions have unequal sizes.

Table 5 covers the radial variations across the Galaxy. Table 5 contains the calculated disk area fraction, which represents the proportion of the surface area of each annular region out of the total disk area (taken out to a radius of 15 kpc, in accordance with the data used).

| Radial Location / kpc | Fraction of Number Density of Stars within this Region: $f_{n(R)}$ | Disk Area Fraction: $\frac{\text{Area of annulus}}{\text{Area of disk}}$ |
|---|---:|---:|
| R < 3 | $0.6991^{+0.0612}_{-0.0538}$ | 0.0400 |
| 3 < R < 5 | $0.1660^{+0.0113}_{-0.0188}$ | 0.0711 |
| 5 < R < 7 | $0.0746^{+0.0144}_{-0.0178}$ | 0.1067 |
| 7 < R < 9 | $0.0335^{+0.0111}_{-0.0116}$ | 0.1422 |
| 9 < R < 11 | $0.0150^{+0.0074}_{-0.0065}$ | 0.1778 |
| 11 < R < 13 | $0.0068^{+0.0044}_{-0.0035}$ | 0.2133 |
| 13 < R < 15 | $0.0030^{+0.0027}_{-0.0018}$ | 0.2489 |
| Totals | 1.0000 | 1.0000 |

Table 5: Displaying the fraction of number density (n) within each radial region of the Milky Way Galaxy (over all vertical positions). In order to generate the fairly-weighted fractions of the absolute number of stars in each region (in Table 11), the disk area fraction will also be required, which is calculated (without associated uncertainties) as follows:

$$\text{Disk area fraction} = \frac{\text{Area of annulus between radii identified}}{\text{Area of MW disk of radius 15 kpc}}$$

Table 6 covers the variation in number density for the vertical variations and takes into account the fraction of the disk thickness which is covered by each region. To calculate this disk thickness fraction, we take the full disk thickness as 4 kpc (2 kpc above and below the midplane) which (given the scale-height we are using, $h_z = 0.335 \, kpc$), represents nearly 6 scale-heights, and hence will encompass approximately 99.8 % of all of the stars, according to the exponential decay model.

| Vertical Distance from midplane / kpc | Fraction of Number Density of Stars within this Region: $f_{n(z)}$ | Disk Thickness Fraction $= \dfrac{Thickness\ of\ region}{4\ kpc}$ |
|---|---|---|
| 0 < \|z\| < 0.5 | $0.7752^{+0.1218}_{-0.1044}$ | 0.25 |
| 0.5 < \|z\| < 1 | $0.1533^{+0.0676}_{-0.0609}$ | 0.25 |
| 1 < \|z\| < 2 | $0.0343^{+0.0622}_{-0.0238}$ | 0.50 |
| \|z\| > 2 | $0.0372^{+0.1891}_{-0.0372}$ | 0.00 |
| Total | 1.0000 | 1.00 |

Table 6: Displaying the fraction of number density (n) within each vertical region of the Milky Way Galaxy (over all radial positions).

As we have seen, the entire Galaxy is divided into 28 annular regions (labelled A to £, see Table 4), and the disk area fraction and disk thickness fraction which each region occupies are used in conjunction with the number density proportions (n) to calculate the fraction of the absolute number of stars (N) which reside within each region, which is shown (together with error bounds) in Table 7.

| z / kpc | R < 3 | 3 < R < 5 | 5 < R < 7 | 7 < R < 9 | 9 < R < 11 | 11 < R < 13 | 13 < R < 15 | TOTALS |
|---|---|---|---|---|---|---|---|---|
| z > 2 | V 0% | W 0% | X 0% | Y 0% | Z 0% | & 0% | £ 0% | 0% |
| 1 < z < 2 | O 1.7% ± 1.7% | P 0.71% ± 0.70% | Q 0.48% ± 0.60% | R 0.29% ± 0.40% | S 0.16% ± 0.30% | T 0.087% ± 0.200% | U 0.045% ± 0.100% | 3.4% ± 3.0% |
| 0.5 < z < 1 | H 3.7% ± 0.6% | I 1.6% ± 0.3% | J 1.1% ± 0.3% | K 0.64% ± 0.30% | L 0.36% ± 0.20% | M 0.19% ± 0.10% | N 0.10% ± 0.10% | 7.7% ± 1.9% |
| -0.5 < z < 0.5 | **A 38% ± 8.1%** | **B 16% ± 3.0%** | **C 11% ± 1.0%** | **D 6.5% ± 0.9%** | **E 3.6% ± 0.9%** | **F 2.0% ± 0.7%** | **G 1.0% ± 0.5%** | **78% ± 15%** |
| -1 < z < -0.5 | H 3.7% ± 0.6% | I 1.6% ± 0.3% | J 1.1% ± 0.3% | K 0.64% ± 0.30% | L 0.36% ± 0.20 | M 0.19% ± 0.10% | N 0.10% ± 0.10% | 7.7% ± 1.9% |
| -2 < z < -1 | O 1.7% ± 1.7% | P 0.71% ± 0.70% | Q 0.48% ± 0.60% | R 0.29% ± 0.40% | S 0.16% ± 0.30% | T 0.087% ± 0.200% | U 0.045% ± 0.100% | 3.4% ± 3.0% |
| z < -2 | V 0% | W 0% | X 0% | Y 0% | Z 0% | & 0% | £ 0% | 0% |
| TOTALS | 49% ± 13% | 21% ± 5.0% | 14% ± 2.8% | 8.4% ± 2.3% | 4.6% ± 1.9% | 2.6% ± 1.3% | 1.3% ± 0.9% | |

Vertical distance from Midplane, z / kpc

Radial Distance from Galactic Centre, R / kpc

Table 7: Relative Fraction of Stars within in each Galactic Region A to £.

The data from Table 4, Tables 13 and 14 in Appendix 1 and Table 7 can be combined to form weighted averages of the fraction of all stars within the Galaxy with metallicities which exceed the three reference values.

This process of creating a weighted average for the fraction of all stars in the Galaxy which exceed the given reference value yields the following results:

Calculated Result (Reference value 1.0 $Z_{solar}$): $f_{M,1.0}$ = $0.4982^{+0.2522}_{-0.1617}$ , therefore: $0.3365 < f_{M,1.0} < 0.7504$

Calculated Result (Reference value 0.5 $Z_{solar}$): $f_{M,0.5}$ = $0.8172^{+0.1828}_{-0.2875}$ , therefore: $0.5297 < f_{M,0.5} < 1.0000$

Calculated Result (Reference value 0.1 $Z_{solar}$): $f_{M,0.1}$ = $0.9738^{+0.0262}_{-0.3723}$ , therefore: $0.6015 < f_{M,0.1} < 1.0000$

Again we find that even in the most pessimistic case in which we assume that the development of life requires a metallicity equal to that of the Sun (which we know has supported life on Earth) we still find that more than one third of stars have sufficient metallicity to potentially support life.

## 3.4 Calculation of $f_{HZ}$: the fraction of stars which host planets in their Circumstellar Habitable Zone.

The habitable zone is an area around a star in which the temperature (with reference to forming `Earth-like' life) is not too high and not too low, and thus life as we know it is able to exist. Other earlier works have also considered the habitable zone as a criterion for finding life around other stars, so in a real sense this assumption is already part of the Astrobiological Copernican principle criteria. Like the number of planets surrounding stars, this quality has long been an unknown, but we are now able to make measurements of it based on work by the Kepler telescope in particular (e.g. Traub et al. 2012; Dressing and Charbonneau 2013).

Traub et al. analyse data from the first 136 days of operation of the Kepler mission, to achieve estimates of the percentage of stars of different spectral types which host planets with particular characteristics. Table 8 of this work, below, presents some key results from that paper: Traub et al. report an average of 0.29 ± 0.02 planets per star for all F, G and K type stars combined; with an average of 0.09 ± 0.01 described as terrestrial planets (with radius between 0.5 and 2.0 Earth radii, corresponding to roughly 0.1 to 10 Earth masses).

Traub et al. then go on to analyse the occurrence of terrestrial planets within the Circumstellar Habitable Zone (HZ) of the host star, and – even if we take the most conservative estimate (corresponding to 0.95 to 1.67AU in our Solar System) – substantial numbers of terrestrial planets are found within the HZ of all F, G and K stars. Dressing and Charbonneau (2013) present similar findings for M-type stars, with the benefit of four years of Kepler data. The most conservative estimate arrived at in this study for the occurrence rate of Earth-sized planets within the HZ for M-type stars is $0.16^{+0.17}_{-0.07}$.

We develop an estimate of $f_{HZ}$ together with its uncertainty in Table 8. We use the overall occurrence rate of terrestrial planets within the HZ of all stars in spectral classes F,G and K, quoted by Traub et al. as 0.34 ± 0.14, together with the value for M-class stars quoted by Dressing and Charbonneau, $0.16^{+0.17}_{-0.07}$. We use this in conjunction with calculated fractions of all stars within each of these categories, which have been derived using the Salpeter IMF (with associated uncertainties, derived from a comparative calculation which employs the Chabrier IMF).

| Spectral Type | Source | Fraction of all Stars in this Spectral Type, based on Salpeter IMF (with uncertainties based on comparative use of Chabrier IMF): (fraction A) | Occurrence rate of terrestrial planets within HZ for this Spectral Type (fraction B) | Component for Weighted Average calculation (Component: AB) |
|---|---|---|---|---|
| FGK | Traub 2012 | $0.145 \pm 0.026$ | $0.34^{+0.14}_{-0.14}$ | $0.0493^{+0.0328}_{-0.0255}$ |
| M | Dressing and Charbonneau 2015 | $0.851 \pm 0.015$ | $0.16^{+0.17}_{-0.07}$ | $0.1362^{+0.1496}_{-0.0610}$ |
| Totals | | $0.996^{+0.004}_{-0.041}$ | N/A | $0.1855^{+0.1824}_{-0.0865}$ |
| Weighted Average Occurrence Rate of Terrestrial Planets around F, G, K and M type stars: | | | | $0.19^{+0.20}_{-0.09}$ |

Table 8: Estimation of $f_{HZ}$ – occurrence rate of terrestrial sized planets within the HZ of all stars within the investigation of this work (FGKM stars) together with its uncertainty.

Hence, we use an average value of $f_{HZ} = 0.19^{+0.20}_{-0.09}$ in the following calculation, i.e. $10\% < f_{HZ} < 39\%$

## 3.5 Summary of Results

Table 9 below presents a summary of the values calculated throughout this work, together with their uncertainties. These values will be utilized in Equation 3, to consider the number and likely spatial distribution of CETI, under each of our modelling assumptions.

| Quantity and Description | | | For use in Category: | Value | Max. % Uncertainty |
|---|---|---|---|---|---|
| $f_{L,weak}$ | fraction of stars with age in range: | age / Gyr > 5.0 (See Section 3.1.3) | Weak 4,5,6 | $0.963^{+0.034}_{-0.183}$ | 19% |
| $f_{L,mod}$ | | 4.0 < age / Gyr < 6.0 Gyr | Moderate 7,8,9 | $0.031^{+0.006}_{-0.006}$ | 19% |
| $f_{L,strong}$ | | 4.5 < age / Gyr < 5.5 Gyr | Strong 10,11,12 | $0.015^{+0.003}_{-0.003}$ | 19% |
| $\tau'_{weak}$ | (Average age of stars / Gyr) – 5 Gyr (See Section 3.2) | | Weak 4,5,6 | $4.80^{+0.55}_{-0.55}$ Gyr | 11% |
| $\tau'_{mod}$ | Time available for life - by definition | | Moderate 7,8,9 | 2.0 Gyr | N/A |
| $\tau'_{strong}$ | Time available for life - by definition | | Strong 10,11,12 | 1.0 Gyr | N/A |
| $f_{M,0.1}$ | fraction of stars with metallicity, Z, exceeding: | $0.1 Z_\odot$ | 1, 4, 7, 10 | $0.9738^{+0.0262}_{-0.3723}$ | 38% |
| $f_{M,0.5}$ | | $0.5 Z_\odot$ | 2, 5, 8, 11 | $0.8172^{+0.1828}_{-0.2875}$ | 35% |
| $f_{M,1.0}$ | (See Section 3.3.2) | $1.0 Z_\odot$ | 3, 6, 9, 12 | $0.4982^{+0.2522}_{-0.1617}$ | 51% |
| $f_{HZ}$ | fraction of F, G, K, M stars with terrestrial planets in Habitable Zone | | All | $0.19^{+0.20}_{-0.09}$ | 110% |

Table 9: Summary of the values calculated throughout Section 3, with their uncertainties.

Note: For our final results, we will also require an estimate of the total number of stars in the Milky Way Galaxy, for which we take the value from Masetti (2015):

$$N_* = 2.5^{+1.5}_{-1.5} \times 10^{11}$$ Maximum Percentage Uncertainty: 60 %

Note that the calculation of the $f_L$ terms is explained in Section 3.1. The uncertainties in these values (19%) are mainly due to the range of values used for the fitting constants in Equation 4, as explained in Table 2, in order to encompass the error bars in the SFR data at different redshifts (Fig 1). This also gives a good error on the uncertainty in the star formation history of the Milky Way which is otherwise unknown.

However, the major source of uncertainty remains in the estimate of $f_{HZ}$ (110%), which will likely improve with increasing data on exoplanet discoveries. Likewise, the range of values used in $N_*$ yield an uncertainty of 60%, which stems from the considerable debate on the distribution of stellar masses throughout the Galaxy.

In the next section, we present our findings based on using the new CETI equation (Equation 3) with the calculated values stated in Table 9.

It will be useful to define the quantity,

$$\kappa = \left(\frac{N_* \cdot f_L \cdot f_{HZ} \cdot f_M}{\tau'}\right)$$

(Equation 19)

such that the number of CETI, $N$, is related to the average lifetime of a civilization, $L$, by:

$$N = \kappa L$$

(Equation 20)

and the values of $\kappa$, for each Modelling Category, are given below, in Table 10, together with the associated uncertainties.

| Category | Comment about A.C.P. | Assumption 1: concerning the time interval available for the existence of life. | | Assumption 2: min stellar metallicity required for CETI. | Number of occurrences of Primitive Life in Galaxy, N |
|---|---|---|---|---|---|
| 1 | Ultra-Weak | (Primitive Life only). Assume that Primitive Life becomes established rapidly, wherever suitable, stable conditions arise, and will persist for the entire stellar lifetime. The fraction $\left(\frac{L}{\tau'}\right)$ in Equation 7 is set to 1, and the term $f_L$ is set to | | $0.1 Z_\odot$ | $4.63^{+11.00}_{-4.02} \times 10^{10}$ |
| 2 | | | | $0.5 Z_\odot$ | $3.88^{+11.70}_{-3.35} \times 10^{10}$ |
| 3 | | | | $1.0 Z_\odot$ | $2.37^{+9.34}_{-2.03} \times 10^{10}$ |
| | | **CETI possible in stellar system of age:** | **Value of $\tau'/Gyr$ implied by Assumption 1.** | | $\kappa / yr^{-1}$ |
| 4 | Weak | (age / Gyr) > 5.0 | (Average star age / Gyr) − (5.0 / Gyr) | $0.1 Z_\odot$ | $9.28^{+19.79}_{-8.18}$ |
| 5 | | | | $0.5 Z_\odot$ | $7.79^{+21.28}_{-6.82}$ |
| 6 | | | | $1.0 Z_\odot$ | $4.75^{+17.07}_{-4.13}$ |
| 7 | Moderate | 4.0 < (age / Gyr) < 6.0 | 2.0 Gyr | $0.1 Z_\odot$ | $0.717^{+2.169}_{-0.642}$ |
| 8 | | | | $0.5 Z_\odot$ | $0.602^{+2.284}_{-0.535}$ |
| 9 | | | | $1.0 Z_\odot$ | $0.367^{+1.799}_{-0.325}$ |
| 10 | Strong | 4.5 < (age / Gyr) < 5.5 | 1.0 Gyr | $0.1 Z_\odot$ | $0.694^{+2.114}_{-0.622}$ |
| 11 | | | | $0.5 Z_\odot$ | $0.582^{+2.226}_{-0.519}$ |
| 12 | | | | $1.0 Z_\odot$ | $0.355^{+1.752}_{-0.315}$ |

Table 10: Describing the twelve categories of differing modelling assumptions, relating to different relative strengths of the Astrobiological Copernican Principle (A.C.P.) Values and uncertainties are shown, for the quantity:

$$\kappa = \left(\frac{N_* . f_L . f_{HZ} . f_M}{\tau'}\right)$$

# 4. The Possible Lifetime and Spatial Distribution of Communicating Intelligent Civilizations in the Galaxy

In this section, we estimate the number and spatial distribution of communicating intelligent civilizations in the Galaxy, based on the assumption that around 5 Gyr is required for the development of such a civilization (i.e. according to the modelling assumptions of the Weak Astrobiological Copernican Principle, Categories 4, 5 and 6).

To sum up, we revisit the terms of the CETI Equation 3. We illustrate our results for the categories Weak 4, 5 and 6 (in which life is assumed to become established any time after 5 Gyr) and we use the values of all of the estimated quantities, as summarised in Section 3.5.

Therefore, we can express Equation 3 such that the number of CETI, $N$, is related to the average lifetime of a civilization, $L$, by Equation 20 (in which we are defining the quantity $\kappa/yr^{-1}$ in Equation 19).

In the Weak Categories (4, 5 and 6), we have:

$$\kappa_4/yr^{-1} = 9.28^{+19.79}_{-8.18}$$

$$\kappa_5/yr^{-1} = 7.79^{+21.28}_{-6.82}$$

$$\kappa_6/yr^{-1} = 4.75^{+17.07}_{-4.13}$$

The simple statement (Equation 20) at least allows a lower limit to be made, given the communicating civilization on Earth has persisted for of order 100 years, which implies that a minimum value for $N$ can be estimated (within our assumptions) by setting $L = 100\,yr$.

$$N_{min} = (100\,yr) \times (\kappa/yr^{-1})$$

(Equation 21)

We can develop the statement into one which deals with the number density of communicating civilizations, $n$, by dividing Equation 20 by the volume of the Galaxy:

$$n/kpc^{-3} = \frac{N}{V_{Galaxy}} = \frac{N}{226} = \frac{\kappa(L/yr)}{226}$$

(Equation 22)

From this, we can make a statement estimating the average volume of space surrounding each communicating civilization, $V_{civ}$.

$$V_{civ} = \frac{1}{n} = \frac{226}{N} = \frac{226}{\kappa L}\ kpc^3$$

(Equation 23)

Taking an approximate value for the thickness of the stellar disk of the Milky Way as 0.3 kpc (see Rix 2013), we can model the volume surrounding each CETI as a cylinder, with a z-dimension of 0.3 kpc, and a radius, $r_{civ}$. This cylindrical volume model will be appropriate as long as the average distance between CETI, $D_{civ} > 0.3\ kpc$: if it is less than this, we will require a model which deals with the spherical volume surround each civilization.

Hence, according to the cylindrical volume model, we can estimate the distance between civilizations, $D_{civ} = 2r_{civ}$, as given by:

$$D_{civ} = 2 \times \sqrt{\frac{V_{civ}}{0.3\ kpc\ \times \pi}\ kpc^3} = 2 \times \sqrt{\frac{226}{N \times 0.3\ kpc\ \times \pi}\ kpc^3} = \sqrt{\frac{960}{N}}\ kpc = 31 \times (N)^{-0.5}\ kpc$$

$$D_{civ} = \frac{31}{\kappa^{0.5}} \times (L)^{-0.5}\ kpc = \frac{101}{\kappa^{0.5}} \times (L)^{-0.5}\ \times 10^3\ light-years$$

(Equation 24)

Once again, the estimate of the minimum value of the lifetime of a CETI civilization, $L > 100\ yr$ (based on our own example) can then be used to express the upper limit on $D_{civ}$. The findings for Weak Category 6 are presented in Table 11. A summary of the values for all twelve modelling categories, including their uncertainties, is given in Table 12.

| Eqn Ref | Calculated Entity | Weak Category 6 |
|---|---|---|
|  | $\kappa/yr^{-1}$ | $4.75^{+17.07}_{-4.13}$ |
| 20 | Expression: Absolute Number of CETI, $N \approx$ | $(4.75^{+17.07}_{-4.13}\ yr) \times \left(\dfrac{L}{yr}\right)$ |
| 21 | Lower limit for $N$, assume $L > 100\ yr$, $N >$ | $475^{+1707}_{-413}$ |
| 22 | Expression: CETI number density, $\dfrac{n}{kpc^{-3}} \approx$ | $(0.021^{+0.076}_{-0.018}\ yr) \times \left(\dfrac{L}{yr}\right)$ |
| 22 | Lower limit for CETI number density, assume L > 100 yr, $\dfrac{n}{kpc^{-3}} >$ | $2.1^{+7.6}_{-1.8}$ |
| 23 | Expression: average volume of space surrounding each CETI, $\dfrac{V_{civ}}{kpc^3} \approx$ | $\dfrac{(47.6^{+318}_{-37.2}\ yr^{-1})}{(L/yr)}$ |
| 23 | Upper limit for average volume of space surrounding each CETI, assume L > 100 yr, $\dfrac{V_{civ}}{kpc^3} <$ | $0.476^{+3.183}_{-0.372}$ |
| 24 | Expression: average distance between CETI, $\dfrac{D_{civ}}{kpc}$ | $(14.2^{+25.2}_{-7.59}\ yr^{-0.5}) \times \left(\dfrac{L}{yr}\right)^{-0.5}$ |
| 24 | Upper limit for average distance between CETI, assume L > 100 yr, $\dfrac{D_{civ}}{kpc} <$ | $1.42^{+2.52}_{-0.76}$ |
| 24 | Expression: average distance between CETI, $\dfrac{D_{civ}}{lightyear}$ | $(46400^{+82200}_{-24700}\ yr^{-0.5}) \times \left(\dfrac{L}{yr}\right)^{-0.5}$ |
| 24 | Upper limit for $D_{civ}$, assume $L > 100\ yr$, $\dfrac{D_{civ}}{lightyear} <$ | $4640^{+8220}_{-2470}$ |

Table 11: Calculated values (based on quantities in Tables 9 and 10) and expressions concerning the spatial distribution of CETI throughout the Galaxy, for Weak Category 6 (See Table 10), in terms of the key unknown parameter, $L$ = the average lifetime of these civilizations, in years.

| | Category | Minimum number in Galaxy, $N >$ | Maximum distance to nearest neighbour, $\dfrac{D_{civ}}{lightyear} <$ | Minimum Expected search time before SETI detects signal / yr |
|---|---|---|---|---|
| 1 | Ultra-Weak (Primitive Life only) | $4.63^{+11.00}_{-4.02} \times 10^{10}$ | $7.36^{+7.17}_{-2.45}$ | N/A |
| 2 | | $3.88^{+11.70}_{-3.35} \times 10^{10}$ | $7.80^{+7.35}_{-2.90}$ | N/A |
| 3 | | $2.37^{+9.34}_{-2.03} \times 10^{10}$ | $9.20^{+8.43}_{-3.80}$ | N/A |
| 4 | Weak | $928^{+1980}_{-818}$ | $3320^{+6300}_{-1440}$ | $1030^{+1070}_{-327}$ |
| 5 | | $779^{+2130}_{-682}$ | $3620^{+6630}_{-1750}$ | $1090^{+1100}_{-389}$ |
| 6 | | $475^{+1710}_{-413}$ | $4640^{+8220}_{-2470}$ | $1290^{+1260}_{-514}$ |
| 7 | Moderate | $72^{+217}_{-64}$ | $11900^{+25000}_{-5990}$ | $2420^{+2720}_{-900}$ |
| 8 | | $60^{+228}_{-54}$ | $13000^{+26000}_{-7080}$ | $2570^{+2770}_{-1050}$ |
| 9 | | $37^{+180}_{-33}$ | $16700^{+32600}_{-9820}$ | $3030^{+3210}_{-1350}$ |
| 10 | Strong | $69^{+211}_{-62}$ | $12100^{+25500}_{-6100}$ | $2450^{+2760}_{-913}$ |
| 11 | | $58^{+223}_{-52}$ | $13200^{+27000}_{-7220}$ | $2600^{+2850}_{-1060}$ |
| 12 | | $36^{+175}_{-32}$ | $17000^{+33600}_{-10000}$ | $3060^{+3280}_{-1370}$ |

Table 12: Summary of key values, together with their uncertainties, to describe the spatial distribution of CETI throughout the Galaxy, according to all Categories of modelling assumptions, and based on the use of 100 years as our value for the lifetime of a typical CETI.

# 5. Discussion

Our findings also provide a fresh perspective on the search for CETI – according to the expressions for the average distance between CETI (from Equation 24) of the form:

$$D_{civ}/lightyear = constant \times L^{-0.5}$$

Figure 5 shows a set of plots of $D_{civ}/lightyear$ vs. $L/years$ for each of the modelling categories (as detailed in Table 1). The points at which the curves cut the diagonal ($D_{civ} = L$) represent the condition that the average lifetime of civilizations is just long enough to make speed of light communication between neighbouring CETI a possibility. In other words, these points on the diagonal allow an estimate of the minimum expected time required before the Search for Extra-terrestrial Intelligence (SETI) yields positive results – these times are recorded in Table 12. For instance, according to the modelling assumptions of Weak Category 4, the minimum expected search time is $1030^{+1070}_{-327}$. If our civilization survives for less than this time, beyond the advent of radio signals which are capable of being detected by neighbouring lifeforms, then it is expected that we will not live long enough to make a positive SETI detection; or - if that civilization mirrors our own - they will not live long enough to receive our return signal. If our survival time can be taken as indicative of the average lifetime of all CETI, then we may imagine a Galaxy in which intelligent life is widespread, but communication unlikely.

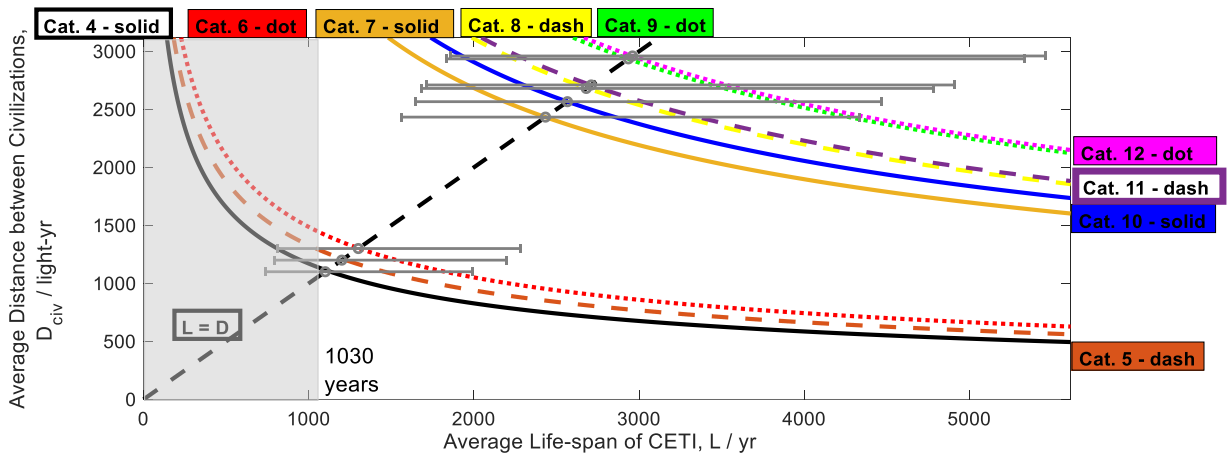

**Fig 5:** Average Distance between CETI, ($D_{civ}$ / light-yr), vs. Average Lifetime of CETI, (L / yr), for Modelling Categories 4 to 12 (See Table 1).

**Diagonal line (L = D) used to determine Minimum Expected SETI Search Time with associated error bars (See Table 15).**

**Shaded region is where average life-span of CETI, L < 1030 yrs: here, species extinction is likely to occur before detection of closest CETI.**

Fig 5 indicates that – in our most optimistic case – we might expect our neighbouring CETI to be approximately 1030 light-years away, therefore the time required for a two-way communication rises to around 2060 years. Indeed, if the average lifetime of civilizations is in fact less than 1030 years, then their average separation becomes too great to allow any communication between neighbours before the species becomes extinct – this scenario is depicted by the grey region in Fig 5. The lifetime of civilizations in our Galaxy is a big unknown within this and is by far the most important factor in the CETI equation we develop, as it was for the Drake equation.

Extinction events are very hard to predict, but they do seem to occur on Earth on a regular basis throughout geological time, due to events such as asteroid collisions. For example, a massive extinction event occurred for dinosaurs after they had existed for 350 million years, but - of course - they were not a communicating intelligence. Part of the issue with our thinking about the lifetime of CETI is that it may be argued (rightly or wrongly) that a civilization's self-destruction is more likely to occur than a natural extinction. Perhaps the key aspect of intelligent life, at least as we know it, is the ability to self-destroy. As far as we can tell, when a civilization develops the technology to communicate over large distances it also has the technology to destroy itself and this is unfortunately likely universal. On Earth, two immediately obvious possibilities are destruction by weapons and through climate change creating an uninhabitable environment. There is, however, another factor that we do not consider here: namely, that the lifetime of an average CETI may be much longer due to space-travel, with civilizations moving off one planet and onto another. This is of course a very difficult thing to do and has not yet been achieved by humans. This would, of course, require that the lifetime of a civilization is long enough such that that event could occur before self-destruction.

Our results also relate in some ways to the so-called Fermi Paradox (i.e. the supposedly surprising failure to detect evidence of extra-terrestrial intelligence after decades of searching) which is often used as an argument against the possibility of the existence of CETI. As detailed by Wright, Kanodia and Lubar (2018), the amount of active SETI carried out to date could hardly be expected to have produced copious positive evidence: they describe the search region as an n-Dimensional Cosmic Haystack (a function of spatial dimensions, time of transmission, sensitivity of receiver, frequency and bandwidth of signal etc) and estimate that active searches thus far have only surveyed a miniscule fraction of this region – some $5.8 \times 10^{-18}$ – which is said to be equivalent to 7700 litres out of the entire Earth's oceans. We may conclude that - whilst a Galaxy-wide CETI in the Milky Way (with an associated large lifetime, L) may be unlikely – CETI with a shorter lifetime is certainly plausible. However, a shorter L would necessarily mean that our closest CETI would be quite distant from Earth and therefore unlikely to be detected for some time, if ever.

# 6. Summary

In this paper we calculate with known uncertainties the number of possible Communicating Extra-Terrestrial Intelligent (CETI) civilizations within our own Galaxy at the present time. We carry out this calculation using the reasonable assumption that life on other planets within the Galaxy develops in broadly similar ways in terms of timescales to life on Earth, although we allow for a range of host star properties and masses. This is the Astrobiological Copernican principle, which asserts that the development of our own intelligent life is not unique or special and similar conditions will produce similar results.

We are able to examine the number of likely communicating advanced civilizations throughout our Galaxy, based on a range of modelling categories (see Table 1). The least strict set of assumptions belong to the Ultra-Weak categories (1, 2 and 3), in which we explore the possibility that primitive life exists wherever stable conditions establish themselves, in the Habitable Zones around stars with sufficient age and metallicity. Such generous assumptions lead to estimated numbers of habitats for primitive life in the Milky Way which reach into the tens of billions.

The main focus of this work, however, is on the possibility of advanced intelligent civilizations, with the ability to communicate over large distances. The Weak categories (4, 5 and 6) are based on the assumption that any suitable habitat which has persisted with stable conditions and adequate chemical richness for at least 5 Gyr should – by comparison with our own example on Earth – have had the same likelihood of developing CETI. The Moderate categories (7, 8 and 9) place this estimate in a tighter framework: namely, we assume that intelligent communicating life can only exist within a 2-billion-year window of opportunity, in stellar systems of age 4 to 6 billion years. The strictest set of scenarios are covered by the Strong categories (10, 11 and 12), in which the window of opportunity for CETI narrows to habitats between the ages of 4.5 and 5.5 billion years.

The starting point in our calculations is the Star Formation Rate (SFR) history, which was at its maximum some 10 Gyr ago. Therefore, we may assume that the peak epoch of life in our Galaxy (and others) would have been around 5 billion years after the peak of the SFR history - which would have been about 5 billion years ago, or at a redshift of $z \sim 0.5$. Hence, our own existence is likely to be somewhat later than the most populous period in Galactic history, assuming CETI has a life-time < 1 Gyr, which could be interpreted as a counterpoint to the Copernican Principle (of the mediocrity of the conditions for our own existence). Indeed, the fact that our Solar System has arrived later than this time of peak formation is intrinsically linked to its high metallicity (since the Sun's parent star must have had sufficient mass to form the heaviest elements in its supernova), so there is potential for future work to explore the concept of apparent anomalies within the Copernican Principle: as time evolves, those systems forming later than the typical time may have a greater propensity for higher metallicity and therefore our own existence – albeit late in time – may still be regarded as a typical occurrence. Cirhovic and Balbi (2019) argue the case for a more subtle redefinition of the concept of temporal typicality, so – from this vantage point – it may be invalid to draw a conclusion about the Copernican Principle from the fact of our own relatively late development.

Of course, the major problem with any speculative analysis of this time has to do with our overwhelming lack of solid evidence for life of a separate lineage to that of the Earth, and the basic premise of suggesting methods of extending our knowledge as to the temporal and spatial distribution of intelligence, based on the single data point that is current available, is indeed a matter for debate. Whilst we have argued that the nature of the development of life on Earth may be used as an exemplar for other systems, other authors (such as Spiegel and Turner, 2011) have used a Bayesian analysis to assert that we cannot necessarily draw such conclusions from the simple fact that Earth's life originated very early in the planet's history.

Overall, we find that in the most limited case, which we describe as the Strong Copernican Astrobiological limit, that there should be a minimum of $36^{+175}_{-32}$ communicating civilizations in the galaxy today, assuming the average lifespan of these civilizations is 100 years. The nearest of these would be at a maximum distance given by $17000^{+33600}_{-10000}$ light-years, making communication or even detection of these systems nearly impossible with present technology. Furthermore, it is almost certain that the host star for this planet host life would be a low mass M-dwarf and not a solar-type star such as our Sun. Indeed, under this strictest set of assumptions, the search for intelligent life is only expected to yield a positive observation if the average life-span of CETI within our Galaxy is $3060^{+3280}_{-1370}$ years (as seen in Table 12, Category 12). That is to say, our communicating civilization here on Earth will need to persist for $6120^{+6560}_{-2740}$ years beyond the advent of long-range radio technology (approximately 100 years ago) before we can expect a SETI two-way communication.

If we relax the assumptions to the Weak Copernican case, we find that there would be a minimum of $928^{+1980}_{-818}$ civilizations communicating in our galaxy today (again, based on a 100 year estimate of average lifetime) with the nearest within a distance of $3320^{+6300}_{-1440}$ light-years away. Under these less strict assumptions, SETI is expected to yield positive findings if the average lifespan of civilizations is $1030^{+1070}_{-327}$ years (as seen in Table 12, Category 4).

Therefore – according to our most limiting set of assumptions and uncertainty bounds – the minimum number of CETI is ~8, with our nearest neighbour at a maximum distance of ~50,000 light-years, which will require ~6300 years of SETI to detect. According to our most generous set of assumptions and uncertainty bounds – the minimum number of CETI is ~2900, with our nearest neighbour at a maximum distance of ~1880 light-years, which will require ~700 years of SETI to detect.

We find that in the much more generous case, in which the lifetime of an average CETI in the Galaxy is a million years – we expect our nearest neighbouring civilization to lie at a distance between 20 and 300 lightyears away. For a perhaps more realistic lifespan of 2000 years we would expect to find a CETI between 400 and 7000 lightyears away. It is clear that the lifetime of a communicating civilization is the key aspect within this problem, and very long lifetimes are needed for those within the Galaxy to contain even a few possible active contemporary civilizations.

If we do not find intelligent life within approximately 7000 lightyears it would indicate one of two things. The first is that the lifetime of civilizations is much shorter than 2000 years, implying that our own may be quite short-lived. The second is that life on Earth is very unique, and intelligent life does not automatically form after 5 Gyr on a suitable planet but is a more random process. It would also imply that intelligences such as 'Life 3.0' artificial life-forms (e.g., Tegmark 2017) created by less robust but intelligent designers (such as ourselves) are unlikely to exist. This type III 'life' can in many ways replicated a 'biological' CETI pattern and are one logical possibility for how a planetary civilization can live for perhaps millions or billions of years without the constraints of 'natural' biological fragility (limited life span, sensitivity to space travel, self-destruction, etc).

The search for intelligent life is therefore a scientific and probabilistic way to determine how long the civilization on Earth is likely to last, or the methods by which life develops. If we do not find life within 10,000 light years for instance, this would be a bad sign for the lifetimes of civilizations, assuming that exo-intelligence is similar to our own or in other words, that the Astrobiological Copernican principle holds.

Acknowledgements
We thank the University of Nottingham School of Physics and Astronomy for its support during the production of this work.

# Appendix 1: Further Data Tables involved in the Calculation of $f_M$

| Distance from Galactic centre, R / kpc | Mean Metallicity / dex | Standard Deviation / dex | Skew-ness (together with its uncertainty) | Calculated Percentages of stars in this region Z < 0.5 $Z_\odot$ |
|---|---|---|---|---|
| Distance from mid-plane, \|z\| / kpc: 0.00 < \|z\| < 0.50 | | | | |
| 3 < R < 5 | +0.23 | 0.24 | −1.68 ± 0.12 | $2.93^{+0.11}_{-0.11}$ % |
| 5 < R < 7 | +0.23 | 0.22 | −1.26 ± 0.08 | $1.47^{+0.04}_{-0.04}$ % |
| **7 < R < 9** | **+0.02** | **0.20** | **−0.53 ± 0.04** | $\mathbf{6.40^{+0.07}_{-0.07}}$ **%** |
| 9 < R < 11 | −0.12 | 0.19 | −0.02 ± 0.03 | $17.12^{+0.12}_{-0.11}$ % |
| 11 < R < 13 | −0.23 | 0.19 | +0.17 ± 0.06 | $34.08^{+0.35}_{-0.37}$ % |
| 13 < R < 15 | −0.43 | 0.18 | +0.47 ± 0.13 | $73.88^{+0.75}_{-0.80}$ % |
| Distance from mid-plane, \|z\| / kpc: 0.50 < \|z\| < 1.00 | | | | |
| 3 < R < 5 | −0.33 | 0.32 | −0.50 ± 0.11 | $58.19^{+0.77}_{-0.84}$ % |
| 5 < R < 7 | −0.18 | 0.29 | −0.50 ± 0.09 | $37.34^{+0.61}_{-0.64}$ % |
| 7 < R < 9 | −0.02 | 0.25 | −0.49 ± 0.06 | $15.31^{+0.24}_{-0.24}$ % |
| 9 < R < 11 | −0.23 | 0.21 | −0.22 ± 0.10 | $38.10^{+0.57}_{-0.59}$ % |
| 11 < R < 13 | −0.27 | 0.19 | +0.28 ± 0.11 | $41.75^{+0.72}_{-0.75}$ % |
| 13 < R < 15 | −0.33 | 0.19 | −0.60 ± 0.39 | $59.10^{+1.42}_{-1.85}$ % |
| Distance from mid-plane, \|z\| / kpc: 1.00 < \|z\| < 2.00 | | | | |
| 3 < R < 5 | −0.27 | 0.29 | −0.48 ± 0.14 | $49.82^{+0.95}_{-1.04}$ % |
| 5 < R < 7 | −0.33 | 0.29 | −0.32 ± 0.13 | $56.77^{+0.94}_{-1.03}$ % |
| 7 < R < 9 | −0.27 | 0.28 | −0.53 ± 0.06 | $49.83^{+0.40}_{-0.40}$ % |
| 9 < R < 11 | −0.27 | 0.25 | −0.37 ± 0.13 | $47.79^{+0.83}_{-0.89}$ % |
| 11 < R < 13 | −0.38 | 0.23 | −0.40 ± 0.21 | $65.88^{+1.03}_{-1.20}$ % |
| 13 < R < 15 | −0.43 | 0.17 | −0.60 ± 0.73 | $79.57^{+1.36}_{-2.52}$ % |

Table 13: Calculated Percentages of stars with metallicity lower than the reference value of 0.5 $Z_\odot$, within each region of the Galaxy. Note: Region D contains the Sun.

| Distance from Galactic centre, R / kpc | Mean Metallicity / dex | Standard Deviation / dex | Skew-ness (together with its uncertainty) | Calculated Percentages of stars in this region with Z < 0.1 $Z_\odot$ |
|---|---|---|---|---|
| Distance from mid-plane, \|z\| / kpc: 0.00 < \|z\| < 0.50 | | | | |
| 3 < R < 5 | +0.23 | 0.24 | −1.68 ± 0.12 | 0.000042% ± 0.000001% |
| 5 < R < 7 | +0.23 | 0.22 | −1.26 ± 0.08 | 0.0000028% ± 0.000001% |
| **7 < R < 9** | **+0.02** | **0.20** | **−0.53 ± 0.04** | **0.000024% ± 0.000001%** |
| 9 < R < 11 | −0.12 | 0.19 | −0.02 ± 0.03 | 0.00018% ± 0.00001% |
| 11 < R < 13 | −0.23 | 0.19 | +0.17 ± 0.06 | 0.0022% ± 0.0001% |
| 13 < R < 15 | −0.43 | 0.18 | +0.47 ± 0.13 | 0.057% ± 0.006% |
| Distance from mid-plane, \|z\| / kpc: 0.50 < \|z\| < 1.00 | | | | |
| 3 < R < 5 | −0.33 | 0.32 | −0.50 ± 0.11 | $2.28^{+0.07}_{-0.09}$ % |
| 5 < R < 7 | −0.18 | 0.29 | −0.50 ± 0.09 | 0.31% ± 0.01% |
| 7 < R < 9 | −0.02 | 0.25 | −0.49 ± 0.06 | 0.0062% ± 0.0002% |
| 9 < R < 11 | −0.23 | 0.21 | −0.22 ± 0.10 | 0.014% ± 0.001% |
| 11 < R < 13 | −0.27 | 0.19 | +0.28 ± 0.11 | 0.0050% ± 0.0004% |
| 13 < R < 15 | −0.33 | 0.19 | −0.60 ± 0.39 | $0.027^{+0.02}_{-0.04}$ % |
| Distance from mid-plane, \|z\| / kpc: 1.00 < \|z\| < 2.00 | | | | |
| 3 < R < 5 | −0.27 | 0.29 | −0.48 ± 0.14 | 0.75% ± 0.04% |
| 5 < R < 7 | −0.33 | 0.29 | −0.32 ± 0.13 | $1.23^{+0.06}_{-0.07}$ % |
| 7 < R < 9 | −0.27 | 0.28 | −0.53 ± 0.06 | 0.59% ± 0.01% |
| 9 < R < 11 | −0.27 | 0.25 | −0.37 ± 0.13 | 0.21% ± 0.01% |
| 11 < R < 13 | −0.38 | 0.23 | −0.40 ± 0.21 | $0.42^{+0.02}_{-0.03}$ % |
| 13 < R < 15 | −0.43 | 0.17 | −0.60 ± 0.73 | $0.049^{+0.003}_{-0.012}$ % |

Table 14: Calculated Percentages of stars with metallicity lower than the reference value of 0.1 $Z_\odot$, within each region of the Galaxy. Note: Region D contains the Sun.